\documentclass[journal=jpcafh,manuscript=article]{achemso}
\usepackage{longtable}
\usepackage{multirow}
\usepackage{amsmath}
\usepackage{subfigure}
\usepackage[usenames,dvipsnames]{xcolor}
\usepackage{soul}
\usepackage{bm}

\usepackage{xr}
\usepackage{amssymb}
\usepackage{siunitx}
\usepackage{threeparttable}
\usepackage{multicol} \usepackage{wrapfig} \usepackage{enumitem}
\usepackage{bm} \usepackage{outlines} 
\usepackage{booktabs}
\usepackage{graphicx}
\usepackage[normalem]{ulem}
\usepackage{hyperref}
\usepackage{lineno}
\usepackage[version=4]{mhchem}
\usepackage[section]{placeins}

\makeatletter
\newcommand*{\addFileDependency}[1]{
  \typeout{(#1)}
  \@addtofilelist{#1}
  \IfFileExists{#1}{}{\typeout{No file #1.}}
}
\makeatother

\newcommand*{\myexternaldocument}[1]{%
    \externaldocument{#1}%
    \addFileDependency{#1.tex}%
    \addFileDependency{#1.aux}%
}
\myexternaldocument{si}

\author{Abhirami Vijayakumar, Raidel Martin Barrios}
\affiliation[University of Basel]{Department of Chemistry, University
  of Basel, Klingelbergstrasse 80, CH-4056 Basel, Switzerland.}

\author{Markus Meuwly} \affiliation[University of Basel]{Department of
  Chemistry, University of Basel, Klingelbergstrasse 80, CH-4056
  Basel, Switzerland.}\email{m.meuwly@unibas.ch}

\title{A State-Space-View of Atom-Diatom Reactions Relevant to
  Rarefied Gas Flow}

\begin{document}
\date{\today}

\begin{abstract}
A microscopically resolved picture of energy flow in atom-diatom
collisions is essential for understanding the non-equilibrium
chemistry in rarefied and hypersonic gas flow. Here, a comprehensive
ensemble of quasi-classical trajectories on global, reactive, and
``vetted'' potential energy surfaces are employed to construct
state-resolved probability maps and to determine the dependence of the
outcomes on the initial ro-vibrational states $(v,j)$. The full range
of processes, including elastic, inelastic, atom exchange, reactive,
and atomization are quantified, revealing distinct structure
reactivity relationships. For the [OOO] system consistent trends are
obtained from two high-quality potential energy surfaces, despite
their different electronic structure and representation
techniques. The resulting state-space description provides a
comprehensive picture of energy redistribution in high-energy
atom-diatom collisions, forming a basis for improved modeling of
non-equilibrium chemistry in hypersonic and rarefied environments.
\end{abstract}

\section{Introduction}
Characterization of state space for polyatomic molecules is at the
heart of chemical physics to address questions related to the flow and
redistribution of vibrational energy, resonance phenomena in
scattering,\cite{nesbitt:1996,wiggins:2010} or the general desire to
uncover and potentially understand the organization and
interrelationship of ``states'' in molecular systems. A particularly
useful model is the ``tier model'' which has been used to describe
intramolecular vibrational energy redistribution (IVR) as a stepwise
diffusion of energy through hierarchically coupled vibrational
states.\cite{gruebele:1998} The initially excited state forms the
zeroth tier, while the first tier contains states directly connected
by first-order anharmonic couplings, and higher tiers include states
reached via successive, weaker interactions. Energy spreads outward
through this network at a rate that is mainly determined by the
coupling strengths between the oscillators. In state-space terms, each
tier corresponds to a collection of quantum states in vibrational
state space. This connection provides a bridge between microscopic
anharmonic couplings and macroscopic energy flow, allowing
quantitative predictions of IVR rates and pathways. Tier models have
also been used in characterizing protein folding to describe the
hierarchical organization of conformational states, where the zeroth
tier represents the native or initial conformation, and higher tiers
consist of states accessible through successive structural
transitions. This framework captures folding kinetics, intermediate
trapping, and pathway heterogeneity to provide a coarse-grained,
simplified, network-based view of the protein’s complex energy
landscape.\cite{frauenfelder:1991,frauenfelder:1996}\\

\noindent
Another context in which the full state space of the chemical species
plays a critical role is rarefied gas flow,\cite{MM.sts:2019,MM.std:2022} a regime encountered at
the high altitudes relevant to hypersonic
flight.\cite{sarma:2000,walpot:2012,dsmc:2017} From a chemical
perspective, the processes occurring in hypersonic flight can be
viewed as a generalization of ordinary burning. While conventional
burning releases energy through fuel oxidation, hypersonic vehicles
experience extreme shock heating that dissociates and ionizes
molecules-primarily N$_2$, O$_2$, and NO-producing radicals and
plasma. Importantly, the chemistry in hypersonic flight is
shock-induced and evolves under extreme non-equilibrium conditions,
with populations far from equilibrium with respect to occupation of
translational, rotational, vibrational, and electronic degrees of
freedom. The energy densities are sufficient to dissociate even small
molecules like N$_2$ and O$_2$, and temperatures in rarefied flow can
reach $T \sim 10000$ K, rising locally to as high as 30000 K.\\

\noindent
Within the broader realm of reaction dynamics for atom–diatom
reactions (A + BC $\rightarrow$ AB + C), the original Polanyi rules
relate the most effective form of reagent energy-translation (T),
vibration (V), and (more weakly and system-dependently) rotation
(R)-to the location of the barrier/transition state along the reaction
coordinate: for an early (reactant-like; in the entrance valley)
transition state, increasing collision/translation energy most
efficiently promotes reactivity, whereas when the barrier is late
(product-like; in the exit valley), vibrational excitation of BC is
typically more effective because it prepares stretched-bond
configurations that resemble those required near the transition state (TS) and improves
coupling into the reaction
coordinate.\cite{polanyi:1969,polanyi:1970}\\

\noindent
Generalizations of the original Polanyi rules increasingly recast
“mode efficacy” for reactions involving polyatomic molecules as a
question of mode-to–reaction-coordinate coupling and energy
redistribution in the TS region, rather than a single early/late
barrier classification.\cite{guo:2016,liu:2023} Product-pair
correlation experiments\cite{liu:2007} and their extension to
“tracking the energy flow along the reaction path”\cite{liu:2008} use
coincident, quantum-state–resolved product pairs to constrain
state-to-state energy partitioning and to infer how particular
reactant preparations correlate with flux through the TS vicinity and
subsequent product energy disposal—thus extending Polanyi-type
intuition into a state-resolved energy-flow picture for
polyatomics. On the theoretical side, the sudden vector projection
(SVP) model provides a complementary TS-local extension: mode
specificity is predicted by the projection of reactant motion (normal
modes or translation) onto the TS reaction-coordinate
(imaginary-frequency) direction at the saddle-point, formalizing which
degrees of freedom couple most directly to barrier crossing in the
sudden limit.\cite{guo:2013,guo:2014}\\

\noindent
Rarefied gas flow in the hypersonic regime is an endeavor of grand
scale.\cite{boyd:2008} Chemical processes, including vibrational
relaxation, atom exchange reactions and full atomization occur on the
picosecond time whereas the time scale on which the object travels is
minutes or hours. Simiarly, the spatial scales range from 1 \AA\/ for
molecular bond lengths to 1 m or more for the object's size. In other
words, hypersonics is inherently a multi-scale problem, spanning a
range of $10^{12}$ in time and space. Notably, all processes occur in
thermal non-equilibrium because hypersonic vehicles travel at speeds
of kilometers per second and experience surface temperatures reaching
and even exceeding 20000 K. The gas flow under such conditions is
dominated by chemistry which is primarily burning of N$_2$, NO, and
O$_2$ and involves all available ro-vibrational states on the reactant
and the product side which leads to $\sim 10^8$ state-to-state cross
sections.\cite{MM.sts:2019,MM.std:2022,MM.std:2022.2}\\

\noindent
The present contribution aims at characterizing the full state space
accessible to atom-diatom collisions including elastic, inelastic,
reactive, and atomization channels. In other words, for a collision A
+ BC$(v,j)$ the outcomes include: (1) A + BC$(v'=v,j'=j)$; (2) A +
BC$(v' \neq v,j' \neq j)$; (3) B + AC$(v',j')$ (3, cyclic
permutations); and (4) A + B + C for which the full state space will
be mapped out and characterized for paradigmatic [ABC] systems
relevant to rarefied gas flow. In addition, the influence of the
underlying potential energy surface (PES) on the final state
distributions will be studied. The systems of interest include the
[OOO], [NOO], and [NNO] collision systems for which accurate,
validated full-dimensional and reactive PESs are available.\\

\noindent
The work is structured as follows. First, the methods employed are
presented, followed by results for the [OOO], [NOO], and [NNO]
collision systems. For each system zone-resolved final state
distributions are discussed and the connection between particular
final states with the initial states they originate from is
investigated. Finally, conclusions are drawn.

\section{Methods}
This section outlines the methodology employed in the present study. A
brief summary of the quasi-classical trajectory protocol and the PESs
used in the present work is provided. In addition, the post-processing
workflow used to analyze the quasi-classical trajectory (QCT) outcomes
is described in detail, including the classification of reaction
channels and the construction of state-to-state distributions required
to quantify the final rovibrational state populations.\\

\subsection{Quasi-classical trajectory simulations}
The QCT method used in this work has been thoroughly described in the
literature.\cite{truhlar:1979,henriksen2008reaction_dynamics,koner:2016,MM.cno:2018,MM.no2:2020}
In this approach, Hamilton’s coupled differential equations of motion
are integrated using the fourth-order Runge--Kutta method in reactant
Jacobi coordinates with a time step of $\Delta t = 0.05$~fs, ensuring
conservation of total energy to within $10^{-6}\,E_\mathrm{h}$ for
each trajectory. After adequate time propagation by monitoring the
internuclear distances, the momenta and positions are transformed to
the appropriate coordinate system (i.e., product Jacobi coordinates if
a product is formed). \\

\noindent
The internal final angular momentum $\mathbf{j}' = \mathbf{q}' \times
\mathbf{p}'$ of the product diatomic species is obtained from the
final position $\mathbf{q}'$ and momentum $\mathbf{p}'$. To determine
the rotational quantum number $j'$ as a real value, the following
quadratic expression is solved\cite{truhlar:1979,karplus:1965}:
\begin{equation}
j' = -\frac{1}{2} 
     + \frac{1}{2} 
       \left( 
         1 + 4\,\frac{\mathbf{j}' \cdot \mathbf{j}'}{\hbar^{2}} 
       \right)^{1/2}.
\label{eq:rotational_quantum}
\end{equation}
The vibrational quantum number ($\nu'$) of the final diatomic species
is subsequently determined as a non-integer according
to\cite{karplus:1965,truhlar:1979}
\begin{equation}
\nu' = -\frac{1}{2} 
       + \frac{1}{\pi \hbar}
         \int_{r^-}^{r^+}
         \left\{
            2\mu \left( E_{\mathrm{int}} - V(r) \right)
            - \frac{\mathbf{j} \cdot \mathbf{j}}{2 m r^{2}}
         \right\}^{1/2}
         \, dr .
\label{eq:vib_quantum}
\end{equation}
where $r$ denotes the diatomic bond length, $r^{+}$ and $r^{-}$
represent the turning points of the diatomic species on the effective
potential corresponding to rotational state $j'$ for internal energy
$E_{\mathrm{int}}$, $\mu$ is the reduced mass of the product
  diatomic, and $V(r)$ is the diatomic potential energy curve,
respectively.\\

\noindent
A total of $2\times 10^{5}$ trajectories was propagated at the
translational energy $E_{\mathrm{trans}} = 3.3607~\mathrm{eV}$ which
is a typical collision energy used in laboratory-based
experiments.\cite{MM.no2:2023}. The impact parameter $b$ was sampled
over the interval $[0,\,b_{\max}]$ with $b_{\max}=10~a_{0}$ using
stratified sampling. Trajectories were integrated up to 75 ps, or
terminated earlier if dissociation/separation criteria were met: a
diatom was considered dissociated when the internuclear distance
exceeded 8 a$_0$ and a new atom-diatom configuration was assigned when
new one internuclear distance was shorter than 8 a$_0$ while the third
atom was more than 16 a$_0$ away from the pair. Initial rovibrational
quantum numbers were sampled to give higher accuracy for low-lying
states: the $(v,j)$ grid was deliberately biased toward small quantum
numbers while retaining coverage of the broader state space. The
vibrational and rotational quantum numbers for all the systems were
sampled from the following unions of arithmetic progressions:
\[
v \in \{0,2,\ldots,12\}\ \cup\ \{15,18,\ldots,30\}\ \cup\ \{34,38,42,46\}, 
\]
\[
j \in \{0,5,10,15\}\ \cup\ \{30,45,60,75\}\ \cup\ \{95,115,\ldots,235\}
\]
For each of the regions defined by these initial conditions the
state-resolved probabilities were determined for each possible
reaction outcome. For the two diatomic molecules in the product
channels, the maximum values of the vibrational and rotational quantum
numbers supported by the 1d-PES for O$_2$ are 49 vibrational and 249
rotational levels, 47 and 240 for NO, and 56 and 251 for N$_2$ respectively.

\subsection{Potential Energy Surfaces and Channels}
QCT trajectories were propagated on global, reactive PESs represented
either as a reproducing-kernel Hilbert space (RKHS)\cite{MM.rkhs:2017}
or as permutationally invariant polynomials (PIPs).\cite{jiang:2016}
The RKHS methodology has been successfully employed to build PESs for
several other triatomic
molecules.\cite{MM.cno:2018,MM.heh2:2019,MM.no2:2020} All RKHS-PESs
used in the present work were generated at the MRCI+Q
\cite{langhoff1974configuration,werner1988efficient} level employing
the augmented correlation-consistent polarized triple-zeta
(aug-cc-pVTZ, avtz) basis set.\cite{dunning1989gaussian} This
methodology has been shown to provide reliable electronic-structure
descriptions for both global and reactive PESs of neutral triatomic
species containing C, N, and
O.\cite{MM.cno:2018,MM.no2:2020,MM.co2:2021} A comprehensive
description of the PES construction is available in
\cite{MM.o3:2025,MM.no2:2020}.\\

\noindent
Each QCT trajectory was broadly assigned to either an elastic, an
inelastic, or a reactive collision or an atomization event. Depending
on the initial states and system considered, there were also
QCT-outcomes that were not assigned to any of the above processes. As
an example, it is possible that for A+BC$\rightarrow$B+AC the
AC-product does not dissociate before the end of the trajectory and
therefore can feature an unphysically large vibrational quantum number
$v'$. A second possibility arises from the specific distance criterion
used to identify products: a diatom is assigned whenever one
internuclear distance is shorter than 8 $a_{0}$, while the third atom
is assumed to be asymptotically separated when its distances exceed 16
$a_{0}$. Under some geometries, two atoms may temporarily satisfy this
condition even though the system is dynamically in the
three-free-atoms channel, leading to a false diatom-atom assignment.\\

\subsection{Initial state sampling and final-state distributions}
QCT simulations were launched over a structured grid of rovibrational
initial states $(v,j)$ of the reagent diatom. To analyze how the
initial preparation controls the outcome, the $(v,j)$ plane was
partitioned into four, non-overlapping zones that isolate distinct
excitation regimes: zone~I corresponds to low internal excitation with
$v\in[0,10]$ and $j\in[0,50]$; zone~II focuses on high rotational
excitations with $v\in[0,10]$ and $j\in[150,250]$; zone~III targets a
balanced, moderately high rovibrational excitation with $v\in[10,30]$
and $j\in[50,150]$; and zone~IV emphasizes initial vibrational
excitation with $v\in[30,50]$ and $j\in[0,50]$. These bounds were
chosen to (i) separate vibration-dominated and rotation-dominated
preparations, (ii) provide a mixed mid-excitation window, and (iii)
retain a low-excitation baseline for comparison. Every initial state
falls into exactly one zone by construction. For each validly
completed QCT-trajectory the output record was classified into exactly
one of the possible product channels together with the final
rovibrational quantum numbers $(v',j')$ following the classification
rules defined above.\\

\noindent
For each product type, a two-dimensional final-state distribution is
reported over $(v',j')$ by considering
\[
P(v',j') \;=\; \frac{N(v',j')}{\sum_{v',j'} N(v',j')},
\]
where $N(v',j')$ is the number of trajectories that terminate in the
final state $(v',j')$. In the implementation, lines of the trajectory
output are parsed to extract $v'$ (x~axis) and $j'$ (y~axis); integer
mapping is performed via rounding to the nearest non-negative quantum
numbers, with entries outside predefined bounds
discarded. Zero-population cells are masked to avoid visual bias, and
a logarithmic color normalization is applied when dynamic range
warrants it. When distributions are reported per zone, the same
normalization is applied within each zone (i.e., each zone’s
$P(v',j')$ integrates to unity over $(v',j')$), enabling shape
comparisons independent of zone population. When a common color scale
is needed across zones, the minimum nonzero and maximum cell values
are pooled to define a shared log scale. These choices mirror the
counting rules used for the bar chart and ensure that (i)
classification is uniform across all panels, (ii) global and
zone-resolved statistics are directly comparable, and (iii) the visual
mapping faithfully reflects the underlying trajectory counts.\\

\section{Results}
\begin{figure}[H]
  \centering
  \includegraphics[width=0.98\textwidth]{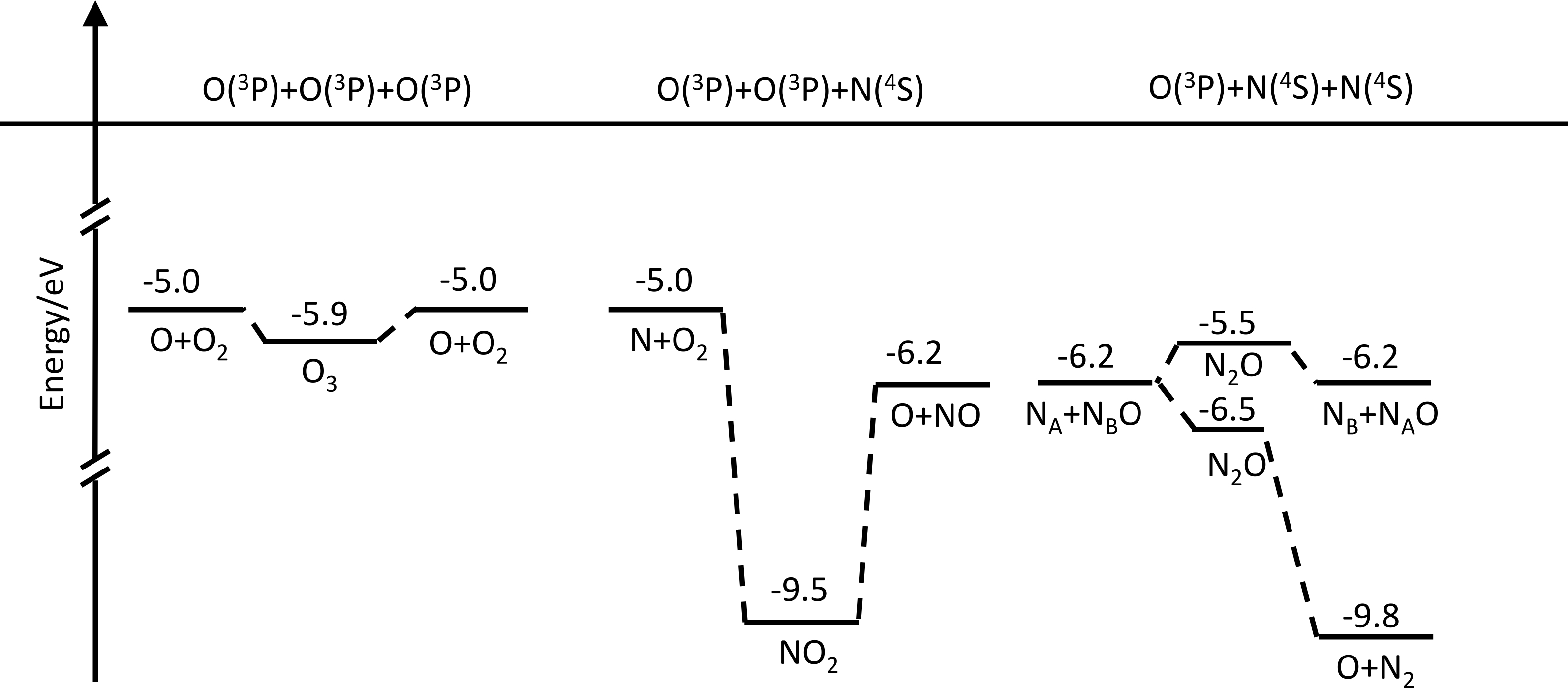}
  \caption{Schematic of all PESs used in the present work; energies
    (in eV) not to scale and no barriers and intermediates
    shown. Panel Left: The O($^3$P) + O$_2$(X$^3\Sigma_{\rm g}^{-})$
    reaction system. Atomization leads to 3O($^3$P). The energies in
    black are from the RKHS-PES.\cite{MM.o3:2025} Middle: The N($^4$S)
    + O$_2$(X$^3 \Sigma^{-}_{\rm g})$ $\rightarrow$ O($^3$P) + NO(X$^2
    \Pi$) reaction.\cite{MM.no2:2020} Right: The N($^4$S) + NO(X$^2
    \Pi$) $\rightarrow$ O($^3$P)+N$_2$(X$^1 \Sigma_{\rm g}^+)$
    collision system.\cite{MM.n2o:2020} For this last reaction the
    true PES involves a considerably larger number of states and
    transition states between them.}
  \label{fig:fig1}
\end{figure}

The results sections provides a comprehensive account of all reaction
channels and final state distributions for particular $(v,j)$ initial
state regions of the diatomic for the O+O$_2$, N+O$_2$, and N+NO
systems in their respective electronic ground states. A brief account
of each reactive PES is given at the beginning of each subsection,
followed by an analysis of the final state distributions. The overall
energetics for all three systems is shown in Figure \ref{fig:fig1}.\\

\subsection{The $\mathrm{O_3}$ System: RKHS-Based PES}
The RKHS-based PES for the O($^3$P) + O$_2$(X$^3\Sigma_g^{-} )$
$\leftrightarrow$ O($^3$P) + O$_2(^3\Sigma_g^{-} )$ system was
constructed at the MRCI+Q level of theory together with the
aug-cc-pVTZ basis set. The MRCI+Q calculations were based on
multistate CASSCF(12,9) calculations and 8 states were included in the
stat-averaged calculations. The grid for the single-channel PESs
included 2269 geometrically feasible ground-state geometries and the
three possible channels (O$_{\rm A}$O$_{\rm B}$+O$_{\rm C}$, O$_{\rm
  A}$O$_{\rm C}$+O$_{\rm B}$, and O$_{\rm B}$O$_{\rm C}$+O$_{\rm A}$)
were mixed using an exponential switching function depending on the
internuclear separations. For the crossing region, a separate data set
was generated. The single-channel PES featured a representation error
of ${\rm RMSD} < 10^{-5}$ eV (0.0002 kcal/mol) and $r^2 = 1.0$ across
9 eV whereas the mixed PES, describing all 3 asymptotic channels,
yielded RMSD = 0.047 eV (1.1 kcal/mol) and $r^2 = 0.9981$ for
``on-grid'', and RMSD = 0.13 eV $(\sim 2.9$ kcal/mol) and $r^2=0.9951$
for "off-grid" points.\cite{MM.o3:2025} The RKHS-PES was thoroughly
validated from comprehensive QCT simulations vis-a-vis thermal rates
$k(T)$ for the atom-exchange and the $T-$dependence of the atomization
reaction.\cite{MM.o3:2025}\\

\begin{figure}[H]
  \centering
  \includegraphics[width=0.98\textwidth]{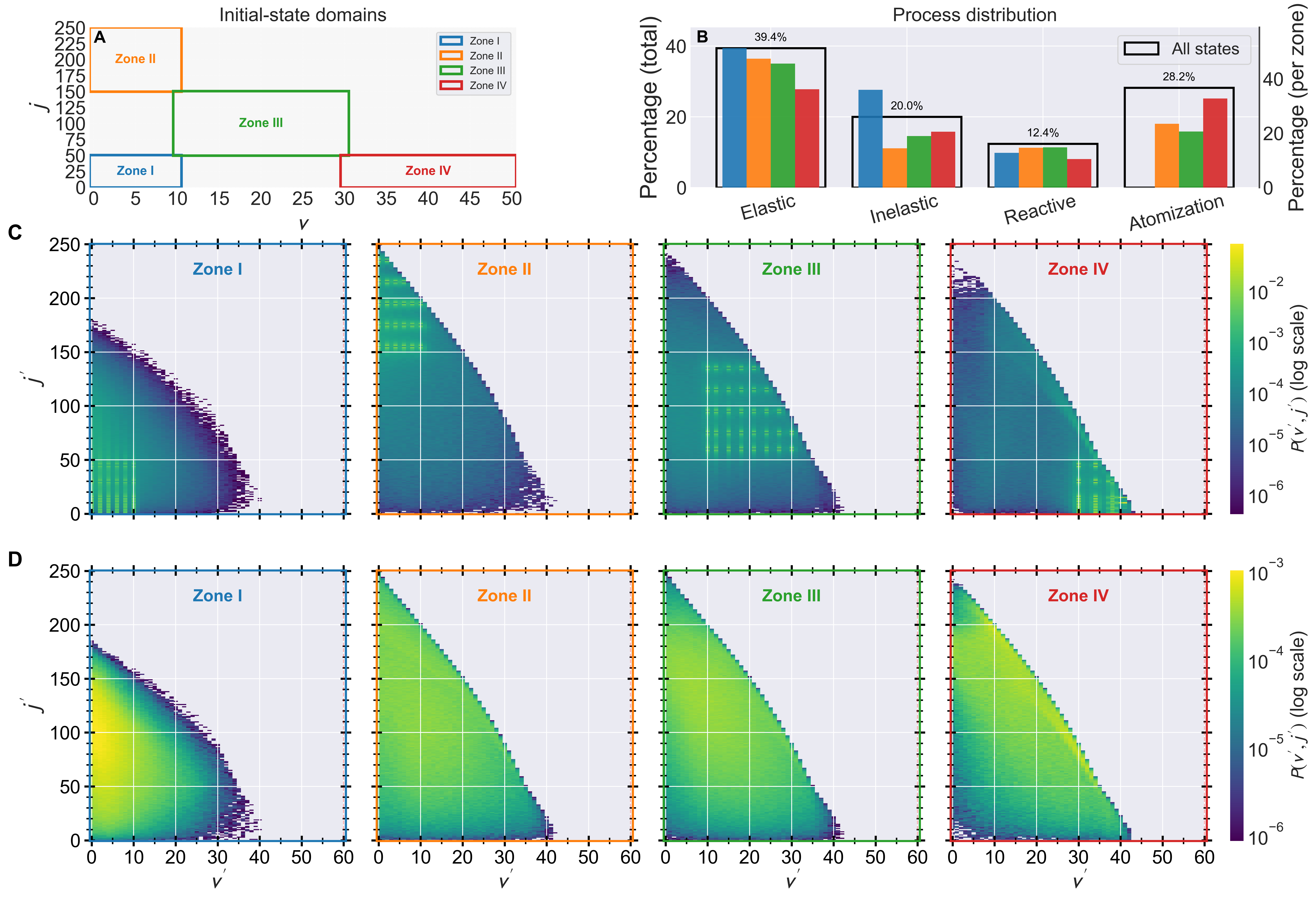}
  \caption{The O($^3$P) + O$_2$(X$^3\Sigma_g^{-} )$ collision
    system. Panel A: Initial-state domains in $(v,j)$ defining
    zones~I--IV: I: $v\in[0,10],\, j\in[0,50]$; II: $v\in[0,10],\,
    j\in[150,250]$; III: $v\in[10,30],\, j\in[50,150]$; IV:
    $v\in[30,50],\, j\in[0,50]$. Panel B: Process distribution: total
    (black outline, no fill) and zone-resolved bars (colored), with
    zone bars normalized to the total so they lie within the outline;
    percentages values are shown only for the total. Panel C:
    Inelastic final-state distributions $P(v',j')$ for zones~I--IV
    (from left to right), respectively. Panel D: Reactive final-state
    distributions $P(v',j')$ for zones~I--IV. Heat maps use a common
    logarithmic color scale within each row. Zone colors are
    consistent across panels, and the zone label appears at the
    top-right of each map. All simulations were carried out using the
    RKHS-PES for O$_3$.}
  \label{fig:fig2}
\end{figure}

\noindent
Figure \ref{fig:fig2} summarizes how the initial rovibrational
preparation $(v,j)$ partitions flux among elastic, inelastic, reactive
(atom exchange), and atomization channels, and how internal energy is
redistributed into product states $(v',j')$. The four initial-state
zones in panel~A were chosen to disentangle the roles of low/high $v$
and $j$. Figure \ref{fig:fig2}B overlays zone-resolved contributions
(bars) on the global process histogram, enabling a direct attribution
of channels to specific initial domains without losing absolute
scale. The black outlined bars quantify the total fraction of each
channel over all $(v,j)$. Colored, inset bars (normalized to the
global total) reveal which initial zones drive each outcome. Zones I
and II with high initial rotation contribute primarily to elastic and
inelastic processes, whereas moderate $(v,j)$ conditions enhance
reactivity and high initial $v$ promotes atomization whereas low$-v$
and low$-j$ prevent atomization, as expected. This establishes that
the appearance and important of a particular mechanism is not uniform
in state space: particular channel yields are selectively gated by the
initial degree of rotational excitation and, to a lesser extent, by
vibrational excitation.\\

\noindent
A rationalization for the comparatively low probability of atomization
trajectories is afforded by Figure \ref{sifig:fig1}. For O$_3$,
atomization requires approximately 5 eV above the O + O$_2$
asymptote. In zone I the diatom is only weakly excited (low $v$ and
$j$), so the total energy $E_{\rm col}$ + $E_{\rm int}(v, j)$ remains
below the three-atom threshold for most initial states. Consequently,
zone I contributes insignificantly to atomization, whereas highly
excited states in zones II–IV can efficiently access the atomization
channel.\\

\noindent
Overall, the dominant pathway is ``elastic scattering'', followed by
``atomization'', with ``inelastic'' and ``reactive'' events occurring
with lower probabilities, see Figure \ref{fig:fig2}B. Zone-resolved
trends reveal clear structure–reactivity relationships. Zone~I
supplies the largest share of inelastic events and a substantial
contribution to elastic scattering. By contrast, reactivity is
promoted primarily by zones II and III, indicating that significant
rotational excitation is conducive to the exchange
pathway. Atomization is favored in zones~II to IV, consistent with the
need for elevated internal energy whether rotational, vibrational, or
a balanced combination to drive complete dissociation. Initial
conditions in zone IV contribute below-average to elastic, inelastic,
and reactive processes but dominate for the atomization
channel. Hence, pronounced initial vibrational excitation together
with a high collision energy promotes full dissociation.\\

\begin{figure}[H]
  \centering
  \includegraphics[width=0.98\textwidth]{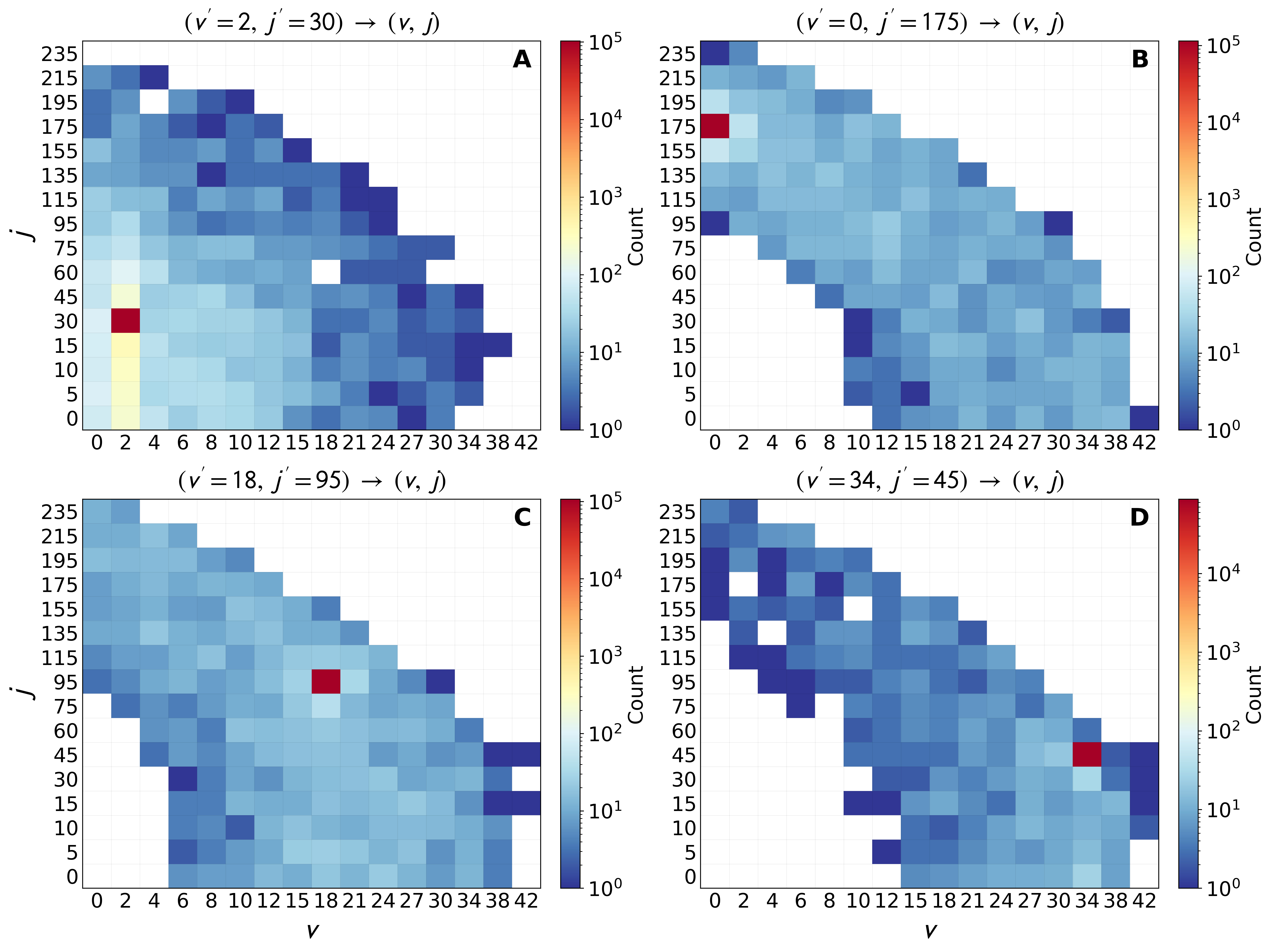}
  \caption{The O($^3$P) + O$_2$(X$^3\Sigma_g^{-} )$ collision
    system. Origin maps of initial rovibrational states $(v,j)$ that
    yield the indicated final states $(v',j')$ from simulations using
    the RKHS-PES for O$_3$. Panels (top–left to bottom–right)
    correspond to $(v',j')=(2,30)$, $(0,175)$, $(18,95)$, and
    $(34,45)$.  Each panel shows a heat map of counts (logarithmic
    scale) of trajectories on the discrete grid of available initial
    states; white cells indicate no events.}
  \label{fig:fig3}
\end{figure}

\noindent
Analysis of the final-state maps, see Figures \ref{fig:fig2}C/D,
further supports these trends: for the reactive channel, initial
conditions from zone~I predominantly populate final states within
zone~I with a propensity for higher $j'-$values. whereas zones~II and
III populate states clustered in the central region of the $(v',j')$
plane and zone~IV tends toward larger $j'$ and smaller $v'$,
consistent with vibrational–to–rotational energy redistribution. For
the inelastic channel, the most populated bins lie close to the
initial-state manifold, indicating that inelastic scattering proceeds
mainly through moderate changes in the rovibrational quantum numbers
($\Delta v,\Delta j$) despite the considerable amount of collisional
energy available.\\

\noindent
A final question concerned the connectivity of a particular final
state with the set of initial states. In other words: how are product
states related to particular initial state preparations, see Figure
\ref{fig:fig3}. For the O+O$_2$ collision this analysis is meaningful
for the elastic, inelastic, and reactive channels. Such ``origin
maps'' were generated for $(v',j') = (2,30); (0,175); (18,95);
(34,45)$ which are part of the set of initial states so that elastic,
inelastic, and reactive processes can be distinguished. The $(v,j)$
axis labels report exactly the set of initial $v-$ and $j-$states and
separations between neighboring states are not necessarily equidistant
($\Delta v = {2,3,4}$ and $\Delta j = {5, 15, 20}$). All probabilities
are reported on a logarithmic scale. The $(v',j') = (0,175), (2, 30),
(18,95), (34,45)$ final states are connected with their origin states
through elastic (44 \% to 57 \%) and inelastic (21 \% to 39 \%)
collisions whereas atom exchange reactions occur in 10 \% to 19 \% of
the cases.  For $(v'=2,j'=30)$ the situation differs in that inelastic
collisions with $(v=2)$ cover a considerably wider range $(0 \leq j
\leq 45)$ of $j-$values.\\

\subsection{The $\mathrm{O_3}$ System: PIP-Based PES}
The PIP-PES for O$_3$ used extended multi-state (XMS) complete active
space second-order perturbation theory (XMS-CASPT2) with a minimally
augmented correlation-consistent polarized valence triple-zeta
(maug-vtz) basis set based on reference states from SA-CASSCF(12,9)
calculations.\cite{varga:2017} In all SA-CASSCF calculations, states
were averaged with dynamical weighting. A level shift of 0.3 E$_h$ was
applied to mitigate intruder state errors. Depending on the energy
range considered (energies $<$ 100, 100-200, 200-500, 500-1000,
0-1000, and $> 1000$) kcal/mol, the RMSEs of the PIP-representation
were 2.9, 4.5, 8.4, 14.5, 6.1, and 26.2 kcal/mol.\\

\begin{figure}[H]
  \centering
  \includegraphics[width=0.98\textwidth]{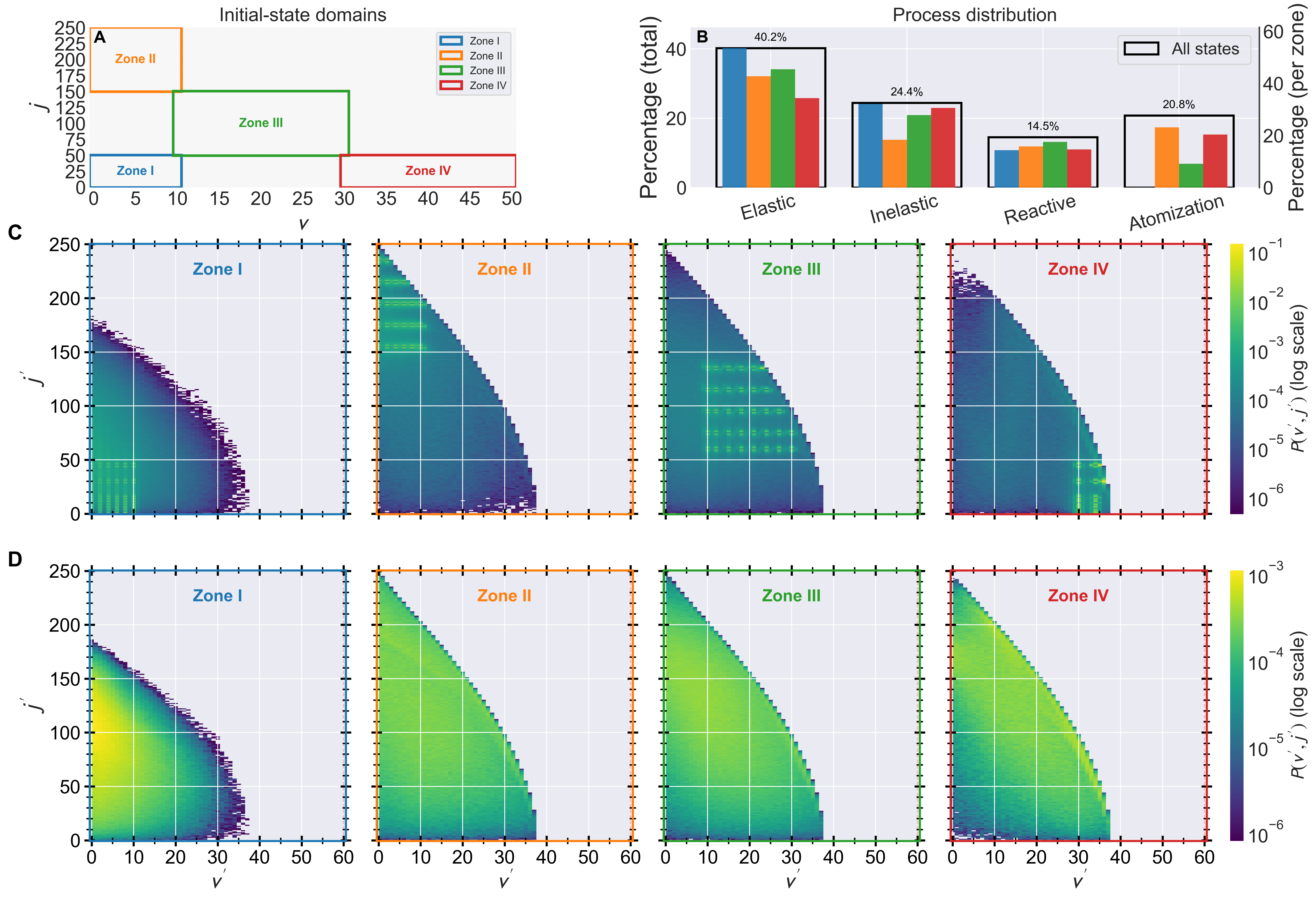}
  \caption{The O($^3$P) + O$_2$(X$^3\Sigma_{\rm g}^{-})$ collision
    system. Panel A: Initial-state domains in $(v,j)$ defining
    zones~I--IV: I: $v\in[0,10],\, j\in[0,50]$; II: $v\in[0,10],\,
    j\in[150,250]$; III: $v\in[10,30],\, j\in[50,150]$; IV:
    $v\in[30,50],\, j\in[0,50]$. Panel B: Process distribution: total
    (black outline, no fill) and zone-resolved bars (colored), with
    zone bars normalized to the total so they lie within the outline;
    percentages values are shown only for the total. Panel C:
    Inelastic final-state distributions $P(v',j')$ for zones~I--IV
    (from left to right), respectively. Panel D: Reactive final-state
    distributions $P(v',j')$ for zones~I--IV. Heat maps use a common
    logarithmic color scale within each row. Zone colors are
    consistent across panels, and the zone label appears at the
    top-right of each map. All simulations were carried out using the
    PIP-PES for O$_3$.}
  \label{fig:fig4}
\end{figure}

\noindent
Figure~\ref{fig:fig4}A shows again zones I to IV in the $(v,j)$ plane
used to classify the initial conditions. Figure \ref{fig:fig4}B
reports the global fraction of trajectories ending in each channel. On
the PIP-PES, the dominant pathway is elastic scattering, followed by
atomization, whereas inelastic and reactive events are less frequent,
which is consistent with the results for the RKHS-PES, see
above. Zone-resolved percentages reveal clear trends: zone~I
contributes most strongly to inelastic scattering and remains relevant
for elastic events, whereas reactivity is promoted primarily by
zones~II and~III (large initial rotational excitation). Atomization is
favored in zones~II--IV, consistent with the need for substantial
internal energy (rotational, vibrational, or a balanced combination)
to drive complete dissociation.\\

\begin{figure}[h!]
    \centering
    \includegraphics[scale=0.4]{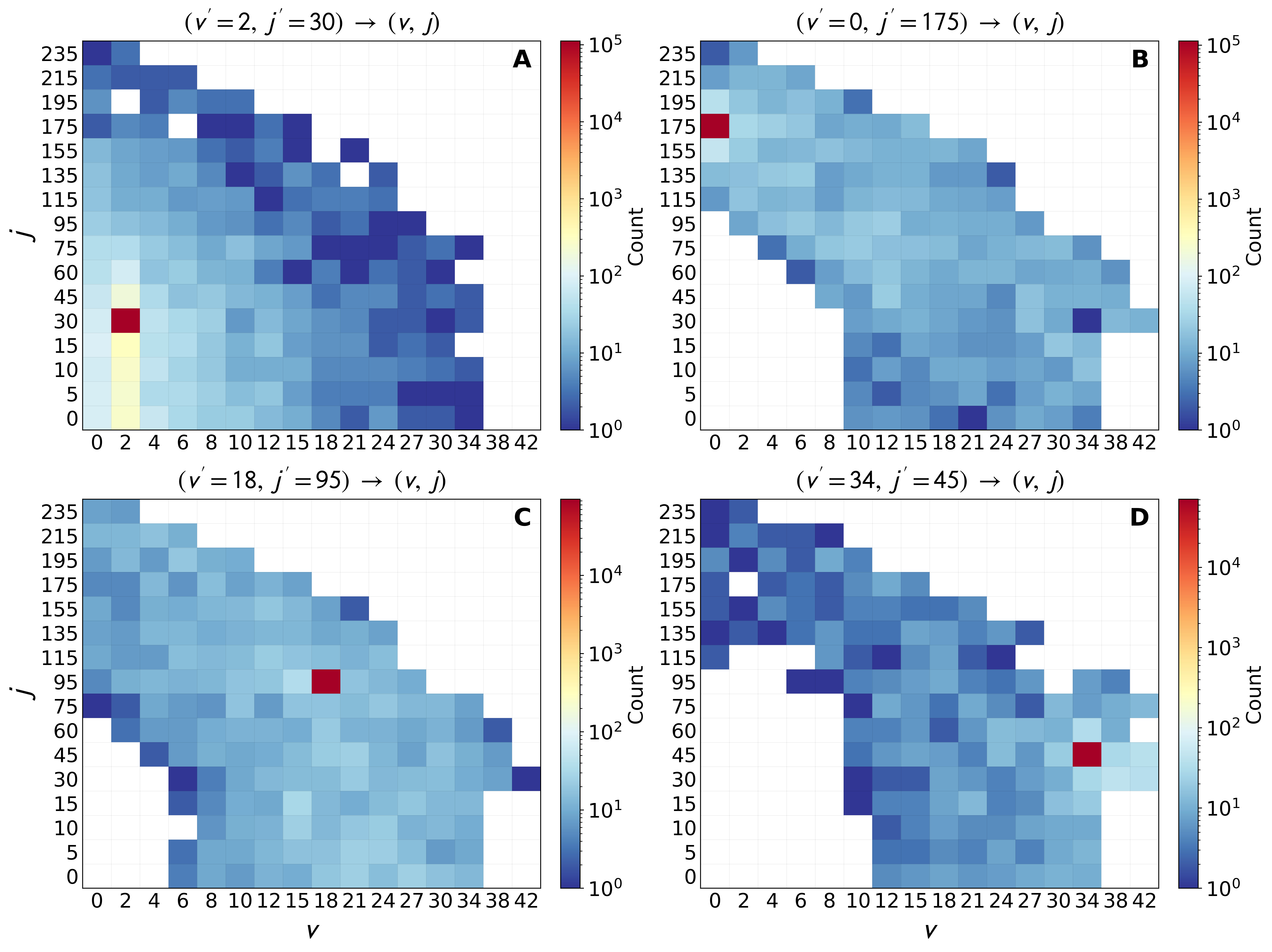}
    \caption{The O($^3$P) + O$_2$(X$^3\Sigma_g^{-} )$ collision
      system. Origin maps of initial rovibrational states $(v,j)$ that
      yield the indicated final states $(v',j')$ from simulations
      using the PIP-PES for O$_3$.\cite{varga:2017} Panels (top–left
      to bottom–right) correspond to $(v',j')=(2,30)$, $(0,175)$,
      $(18,95)$, and $(34,45)$.  Each panel shows a heat map of counts
      (logarithmic scale) of trajectories on the discrete grid of
      available initial states; white cells indicate no events.}
    \label{fig:fig5}
\end{figure}

\noindent
The final-state maps $P(v',j')$, in Figures \ref{fig:fig4}C/D
highlight how internal energy is redistributed on the PIP-PES.  In the
inelastic row (panel C), the most populated bins lie \emph{near} the
initial-state manifold in $(v',j')$, confirming that inelastic
scattering proceeds mainly through small changes in quantum numbers
(modest $\Delta v$ and $\Delta j$). In the reactive row (panel D),
zone~I populates low-$v'$, low-$j'$ products with a slight drift
toward higher $j'$, while zones~II and~III concentrate probability in
the central region of the $(v',j')$ plane; zone~IV tends to yield
products with higher $j'$ and lower $v'$, indicating
vibrational-to-rotational energy transfer along the exchange
pathway. Overall, the PIP-PES reproduces the findings from analysis of
the QCT simulations using the RKHS-PES observed in Figure
\ref{fig:fig2} while modulating the absolute channel weights and the
detailed structure of the product-state distributions. Finally, the
origin map, shown in Figure \ref{fig:fig5}, shares most features with
those from the QCT simulations based on the RKHS-PES, confirming that
the two PESs yield comparable results despite their very different
origins. Even the relative probabilities to form $(v',j') = (0,175),
(2, 30), (18,95), (34,45)$ final states are comparable. For the
elastic, inelastic and atom exchange channels they are (35 \% to 57
\%), (20 \% to 34 \%), and (12 \% to 20 \%).\\

\noindent
It is useful to mention that earlier QCT simulations using the RKHS-
and PIP-based PESs - based on rather different levels of quantum
chemical theory (MRCI+Q/avtz vs. CASPT2/maug-vtz) - yield thermal
rates for the atom exchange and atomization reactions which are both,
consistent with one another and with experiments.\cite{MM.o3:2025}
Also, trained state-to-distribution models for final states are
comparable. This is mirrored in the present findings that the process
distributions (Figures \ref{fig:fig2}B and \ref{fig:fig4}B) follow the
same trends and support the notion that both PESs are of comparable
quality and equally suited to describe the reaction dynamics of the
O+O$_2$ system.\\

\subsection{The N+O$_2$ System}
Reference calculations for the $^2$A$'$ state of NO$_2$ connecting the
N($^4$S) + O$_2$(X$^3\Sigma^-_{\rm g})$ $\rightarrow$ O($^3$P) +
NO(X$^2\Pi)$ states had been carried out at the MRCI+Q/avtz level of
theory, based on CASSCF calculations including all valence orbitals in
the active space. For the O+NO and N+O$_2$ channels the grid contained
7280 and 3920 geometries, respectively. As for the RKHS-PES for the
O+O$_2$ system, the individual channels were mixed using an
exponential switching which employed a separate mixing data set. For
On- and off-grid points the $r^2-$values were above 0.99 with
corresponding RMSE-values of 0.02 eV (0.5 kcal/mol) and 0.03 eV (0.7
kcal/mol).\cite{MM.no2:2020} Compared to the O+O$_2$ collision system,
the N+O$_2$ system is also symmetric in the products of a reactive
trajectory in that both final states, N+O$_2$ $\rightarrow$ O$_{\rm
  A}$ + NO$_{\rm B}$ and $\rightarrow$ O$_{\rm B}$ + NO$_{\rm A}$, are
indistinguishable. Nevertheless, the symmetry is reduced compared with
the O+O$_2$ system discussed so far. This "vetted" RKHS-PES correctly
describes thermal rates for the forward and reverse reactions, the
$T-$dependence of the equilibrium constant, and the $T-$dependence of
the vibrational relaxation for the O+NO collision.\cite{MM.no2:2020}\\

\begin{figure}[h!]
    \centering
    \includegraphics[width=1.0\linewidth]{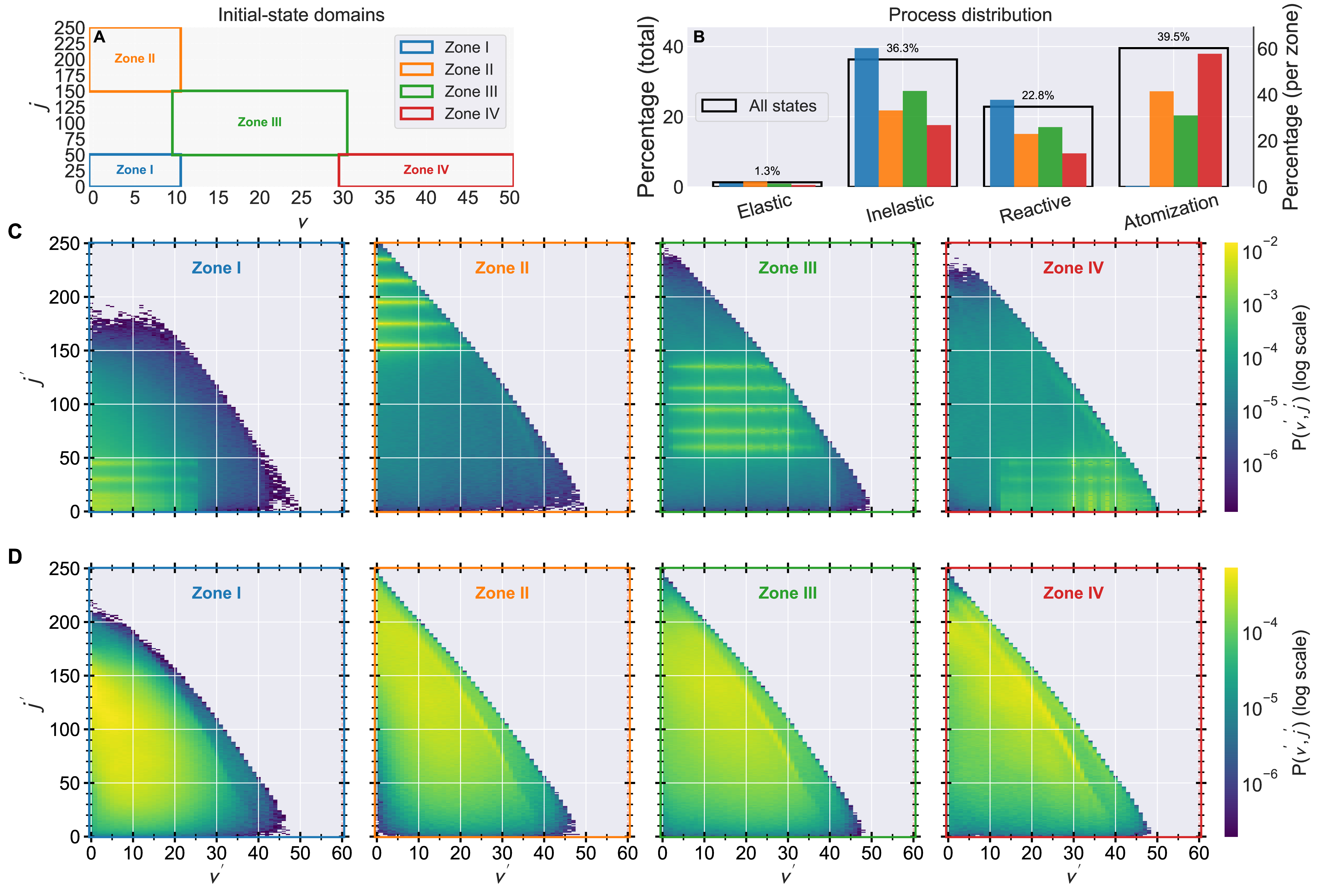}
    \caption{The N($^4$S) + O$_2$(X$^3\Sigma^-_{\rm g})$ $\rightarrow$
      O($^3$P) + NO(X$^2\Pi)$ collision system. Panel A: Initial
      $(v,j)$ domains defining zones~I--IV: I: $v\in[0,10],
      j\in[0,50]$; II: $v\in[0,10], j\in[150,250]$; III: $v\in[10,30],
      j\in[50,150]$; IV: $v\in[30,50], j\in[0,50]$. Panel B: Process
      distribution showing total (black outline) and zone-resolved
      (colored) bars, normalized to the total; percentages shown only
      for the total. Panels C and D report inelastic and reactive
      final-state distributions $P(v',j')$ for zones~I--IV,
      respectively. Heat maps share logarithmic color normalization
      per row. Zone colors and labels are consistent across
      panels. All simulations were carried out with the RKHS-PES for
      NO$_2$.}
    \label{fig:fig6}
\end{figure}

\noindent
Figure \ref{fig:fig6}A presents the initial state domains, the overall
distributions for the different final states (panel B), and the maps
to the final $(v',j')$ states of the NO-products for reactions (panel
C) and inelastic collisions (panel D). Overall, the
N($^4$S)+O$_2$(X$^3 \Sigma_{\rm g}^+$) $\rightarrow$ NO$_2$(X$^2$A')
$\rightarrow$ O($^3$P) + NO($^2 \Pi$) reaction is exothermic (see
Figure \ref{fig:fig1}) with a barrier in the entrance channel (see
Figure S3 in Ref.\cite{MM.no2:2020}). This partly explains the small
probability for elastic processes ($\sim 1 \%$ for all 4 zones). The
most probable channel is decay into atomic fragments, followed by
inelastic and reactive processes. For zone I no atomization is found
whereas this is the most likely channel for zone IV.\\

\begin{figure}[h!]
    \centering
    \includegraphics[width=1.0\linewidth]{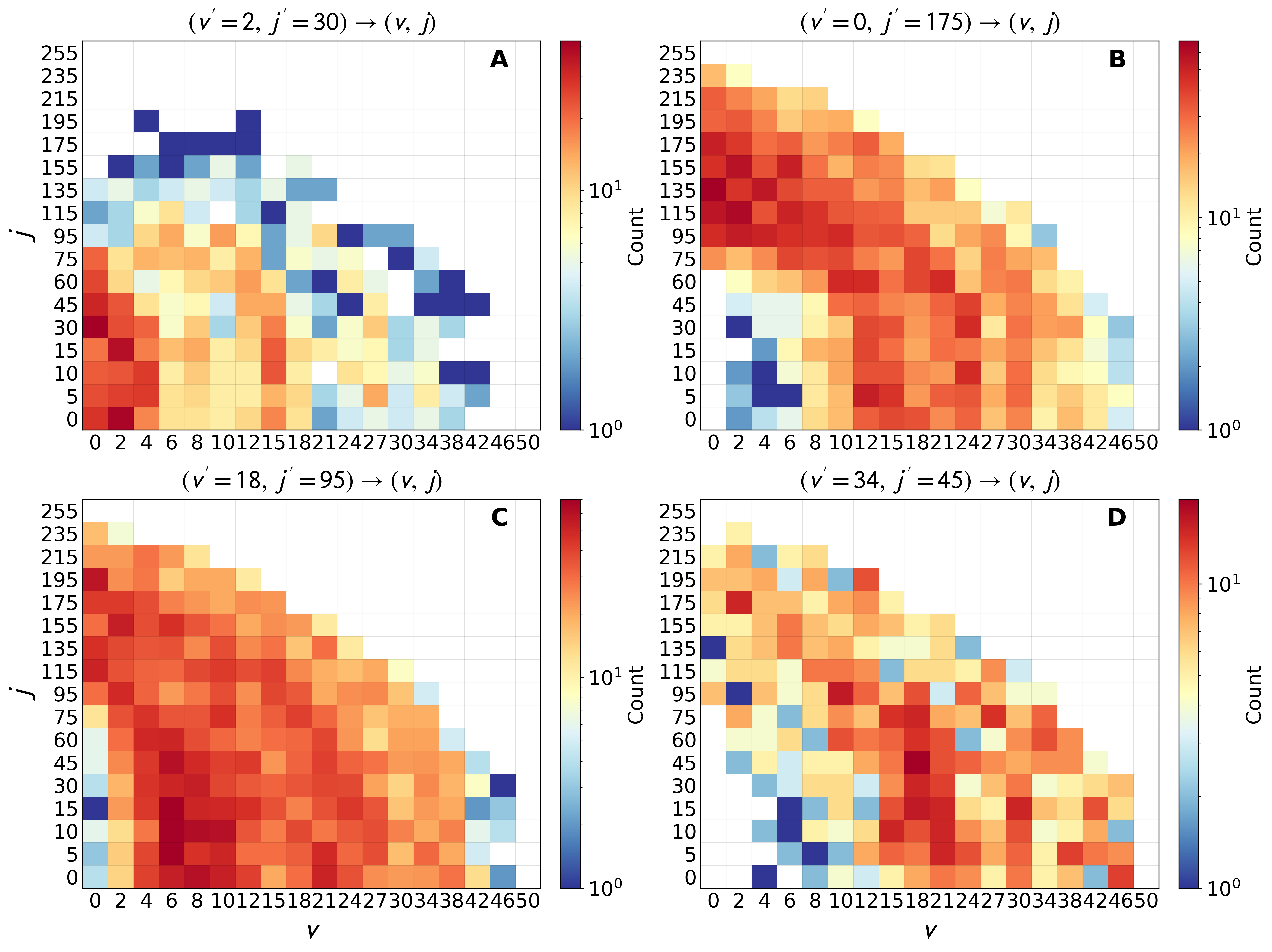}
    \caption{The N($^4$S)+O$_2$(X$^3 \Sigma_{\rm g}^+$) collision
      system. Origin maps for initial rovibrational states O$_2 (v,j)$
      that yield the indicated final states NO$(v',j')$ for the
      reactive channel N($^4$S) + O$_2$(X$^3\Sigma^-_{\rm g})$
      $\rightarrow$ O($^3$P) + NO(X$^2\Pi)$. Panels (top–left to
      bottom–right) correspond to $(v',j')=(2,30)$, $(0,175)$,
      $(18,95)$, and $(34,45)$. Each panel shows a heat map of
      log-scaled counts of trajectories on the discrete grid of
      available initial states; white cells indicate no events. QCT simulations were carried out with the
      RKHS-PES.\cite{MM.no2:2020}}
    \label{fig:fig7}
\end{figure}

\noindent
At the collision energy used, initial states for each of the zones (I
to IV) map onto NO$(v',j')$ states with comparable boundaries although
for zone I the final states extend further towards larger $j'-$values
than for the other zones. This may be a reason for the increased
probability for this reaction channel for zone I compared with all
other zones. On an absolute scale, the probability for a reactive
trajectory is a factor of $\sim 2$ smaller than for inelastic
collisions which makes this channel second-smallest (23 \% overall
probability). It is noted that for reactive trajectories the final
state space is almost fully covered independent of the zone in which
the initial condition is located.\\

\noindent
Based on the definition of the four zones, the origin of final
NO$(v',j')$ (reactive trajectories) and O$_2 (v',j')$ (nonreactive
trajectories) states with $(v',j')=(2,30), (0,175), (18,95), (34,45)$
was analyzed separately in Figures \ref{fig:fig7} and
\ref{sifig:fig2}. For the reactive trajectories that
generate NO$(v',j')$, the NO$(v'=2,j'=30)$ final state originates
primarily from O$_2 (v,j)-$ states corresponding to zone I. This is
not so for the other three final states. Final NO$(v'=0,j'=175)$
originates from both O$_2$ low$-v$/high$-j$ and moderate$-v$/low$-j$
initial states (Figure \ref{fig:fig7}B). The NO$(v'=18,j'=95)$
final state is primarily linked with low$-v$/all$-j$ states from
O$_2$, whereas NO$(v'=34,j'=45)$ is generated from
moderate$-v$/moderate$-j$ states of O$_2$, see Figures
\ref{fig:fig7}C/D. \\

\noindent
For the N($^4$S) + O$_2$(X$^3\Sigma^-_{\rm g})$ $\leftrightarrow$
NO$_2$ process the origin maps include elastic and inelastic
processes, see Figure \ref{sifig:fig2}. At the
state-to-state level, the elastic process is always most likely but
neighboring inelastic channels can be of comparable importance and
clearly outnumber the single elastic process. the widest distribution
of inelastic processes is found for O$_2 (v'=2,j'=30)$, followed by
O$_2 (v'=18,j'=95)$.\\

\subsection{The N+NO System}
For the [NNO] collision system the reference MRCI+Q/avtz calculations
for the two asymptotic channels (O+N$_2$ and N+NO) were carried out on
two grids to account for the different shapes of the PESs in the
asymptotic region. The CASSCF calculations had all $1s$ orbitals
'closed' with the $2s$ and $2p$ orbitals of the nitrogen and oxygen
atoms 'active'. A total of $\sim 12000$ reference energies were used
for the RKHS-representation of the reactive PES. The final performance
for the $^3$A$'$ state featured RMSE = 0.022 eV (0.51 kcal/mol) and
$r^2 = 0.99995$ for on-grid and RMSE = 0.036 eV (0.82 kcal/mol) and
$r^2 = 0.99985$ for off-grid points.\cite{MM.n2o:2020} The highly
exothermic reaction (N($^4$S) + NO(X$^2 \Pi$) $\rightarrow$
N$_2$O(X$^3$A$'$) $\rightarrow$ O($^3$P)+N$_2$(X$^1 \Sigma_{\rm
  g}^+$)) proceeds across two different transition states (TSs) 9.58
kcal/mol and 32.93 kcal/mol above the N+NO asymptote. The reactive
channel bifurcates in that formation of the N$_2$ product proceeds
over the lower-lying transition state whereas exchange of the nitrogen
atoms (O$_{\rm A}$ + NO$_{\rm B}$ $\rightarrow$ O$_{\rm B}$ + NO$_{\rm
  A}$) involves the higher-lying transition state. In contrast to the
N+O$_2$ reaction, the N$_2$O triatomic is not the lowest energy
structure along the reaction pathway. Furthermore, 3 different [NNO]
local minima with different geometries exist.\cite{MM.n2o:2020} QCT
simulations using this RKHS-PES yield good agreement with measured
properties, including thermal rates for the forward and reverse
reactions, the atomization reaction, and vibrational
relaxation.\cite{MM.n2o:2020} The PES was also successful to capture
the low-$T$ kinetics after suitable fine-tuning.\cite{MM.n2o:2023} \\

\begin{figure}[h!]
  \centering
  \includegraphics[width=0.98\textwidth]{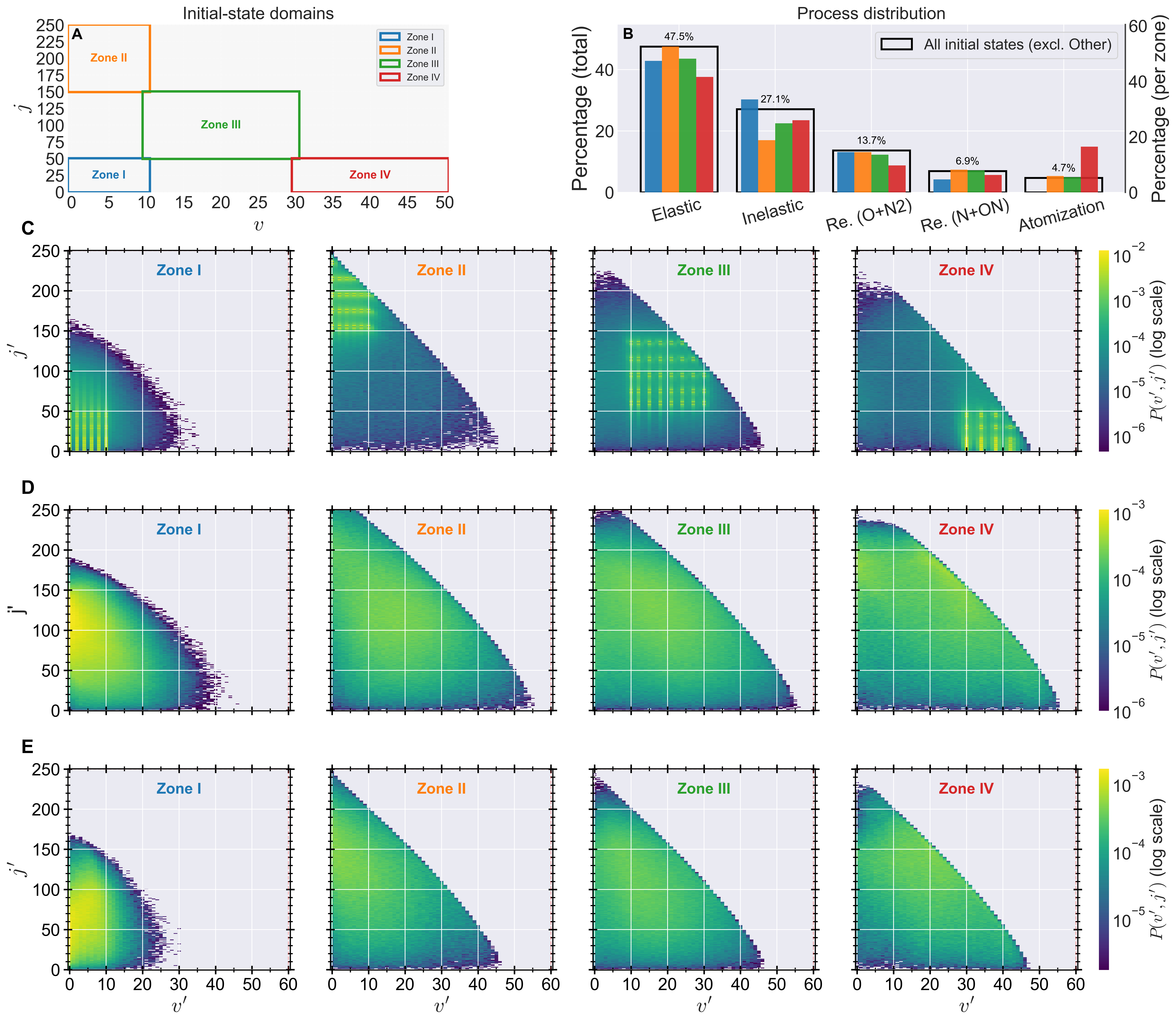}
  \caption{The N($^4$S) + NO(X$^2 \Pi$) collision system. Panel A:
    Initial-state domains in $(v,j)$ defining zones~I--IV: I:
    $v\in[0,10],\, j\in[0,50]$; II: $v\in[0,10],\, j\in[150,250]$;
    III: $v\in[10,30],\, j\in[50,150]$; IV: $v\in[30,50],\,
    j\in[0,50]$. Panel B: Process distribution: total (black outline,
    no fill) and zone-resolved bars (colored), with zone bars
    normalized to the total so they lie within the outline;
    percentages values are shown only for the total. Panels C to E:
    Inelastic, reactive (O+N$_2$), and atom exchange (N+NO)
    final-state distributions $P(v',j')$ for zones~I--IV,
    respectively.  Heat maps use a common logarithmic color
    normalization within each row.  Zone colors are consistent across
    panels, and the zone label appears at the top-right of each
    map. All simulations were carried out using the RKHS-PES for
    N$_2$O.}
  \label{fig:fig8}
\end{figure}

\noindent
Figure \ref{fig:fig8}B shows that the most likely product state
channel are inelastic processes, followed by elastic, reactive, and
atomization. Among the two possible reactions, N$_2-$formation is
favoured by a factor of 2 over N-atom exchange. For the
N$_2-$production reaction (panel C), initial states in zone I lead to
pronounced rotational excitation of the products. This must be due to
the angular anisotropy of the PES. Initial conditions in zones II to
IV do not show particularly pronounced differences except for the fact
that moderately high N$_2 (v',j')-$values for zone IV are more
prevalent than for zones II and III. The N-atom-exchange channel
behaves rather differently. Initial conditions from zone I map rather
closely on the same final-state space with the particularity that two
families of $(v',j')-$states emerge. They are characterized by $v'
\sim 0$ and $v' \sim 10$. For zones II to IV the final state maps are
comparable with the exception that zone II is unlikely to yield $j'
\sim 0$ for the NO product. For inelastic processes the observations
are comparable to the N+O$_2$ reaction in that the boundaries of the
initial and final state spaces are rather congruent. It is noteworthy
that atomization from initial conditions in zone I are not observed at
all whereas zone IV leads to the largest fraction of such
trajectories.\\

\noindent
Origin maps for the N($^4$S) + NO(X$^2 \Pi$) $\rightarrow$
O($^3$P)+N$_2$(X$^1 \Sigma_{\rm g}^+$) and N($^4$S) + NO(X$^2 \Pi$)
$\rightarrow$ N($^4$S) + NO(X$^2 \Pi$) for $(v',j') = (2,30)$;
(0,175); (18,95); (34,45) are reported in Figure
\ref{sifig:fig3}. Each panel shows a heat map of log-scaled counts of
trajectories on the discrete grid of available initial
states. Interestingly, the location of the maxima in the reverse maps
is reminiscent to that of Figure \ref{fig:fig3}. This can be traced
back to the similar relative contributions of the elastic, inelastic,
and reactive channels in both datasets, which involve comparable
statistical weights on the accessible regions of the $(v,j)$
space. The elastic channel imposes the strongest statistical imprint
on the maps, producing sharp maxima at $(v'=v,j'=j)$ regardless of the
dynamical details of competing inelastic or reactive processes.\\

\section{Conclusions}
The present work investigates the final state space available for
three paradigmatic reactions relevant to rarefied gas flow as it
occurs in hypersonics. The collision energies are invariably high and
all possible fragment channels, including elastic, inelastic,
atom-exchange, reaction and atomization processes are possible. For
O$_3$, two competitively accurate and global PESs based on different
levels of quantum chemical theory and their representation, are
available. Both PESs yield accurate thermal rates for the atom
exchange rate and the correct temperature dependence for atomization
when compared with measurements. In that sense, the RKHS- and PIP-PESs
are ``vetted''. It is demonstrated that both PESs yield consistent
results even at the state-resolved level which is remarkable and
reassuring.\\

\noindent
Despite their wide usefulness of the Polanyi-rules, these qualitative
rules are well known to require refinement for complex-forming
dynamics, strong mode-specific coupling, or multi-pathway (including
non-statistical) behavior. Reactions in rarefied gas flow are one such
class of reactions and the present work provides a broad overview of
the state space involved for the [OOO], [NOO], and [NNO] systems. Even
application of product pair correlation or SVP models need to be
critically assessed as they make their own assumptions. For example,
the validity of the pair correlation model can hinge on whether IVR
and mode-mode coupling are significant on the time scale the system
spends in the TS region. Incomplete IVR and dynamical “memory” can
preserve mode specificity and complicate the interpretation due to
non-equilibrium effects. On the other hand, the quantitative
usefulness of the SVP model needs to be assessed case-by-case when the
TS region is strongly anharmonic, exhibits substantial mode
mixing/curvature coupling, or when indirect and nonequilibrium
dynamics undermine the “sudden” assumption.\\

\noindent
One important next step is to map these scenarios on the underlying
PESs to link the actual dynamics with the energies and forces
governing it. This has been done in the past for selective processes,
for example for the [OOO] collision system. Here it has been found
that atom-exchange reactions are characterized by a direct attack of
the O--O bond by the incoming O-atom ($\theta \sim 90^\circ$) whereas
atomization trajectories predominantly approach the diatomic along the
O--O bond, i.e. $\theta = 0$ or $\theta = 180^\circ$. Generalizing
this for all other systems may provide a deeper understanding of the
relationship between the shape of the PES and the reaction
outcomes. Such questions can now be addressed as sufficiently
high-quality, global and reactive PES are available and statistically
significant numbers of QCT simulations can be carried out. It should,
however, be noted, that some of the initial state maps reported in the
present work are probably not yet fully converged, in particular for
the low-probability events, despite the extensive sampling already
used.\\

\noindent
In summary, the present work provides an overview of the accessible
channels, their probabilities, and state-resolved initial- and
final-state embeddings in the full state space for paradigmatic
atom-diatom reactions relevant to hypersonics. Given that the data for
such a characterization can be converged and analyzed in targeted
ways, such an approach opens up the possibility for a data-driven
approach (machine learning) to detecting ``rules'' that govern
particular reaction types. \\

\section*{Supplementary Material}
The supplementary material reports a ro-vibrational energy map for
O$_3$, and additional origin maps for the three reactions.

\section*{Data Availability}
The codes and data for the present study are available from
\url{https://github.com/MMunibas/finalstates} upon publication.

\section*{Acknowledgment}
The authors gratefully acknowledge financial support from the Swiss
National Science Foundation through grants $200020\_219779$ (MM),
$200021\_215088$ (MM), and the University of Basel (MM). This article
is also based upon work within COST Action COSY CA21101, supported by
COST (European Cooperation in Science and Technology) (to MM).\\

\bibliography{bib}

\clearpage
\appendix

\renewcommand{\thefigure}{S\arabic{figure}}
\setcounter{figure}{0}
\section*{Supporting Information: A State-Space-View of Atom-Diatom
	Reactions Relevant to Rarefied Gas Flow}

\begin{figure}[h!]
	\centering
	\includegraphics[width=1.0\linewidth]{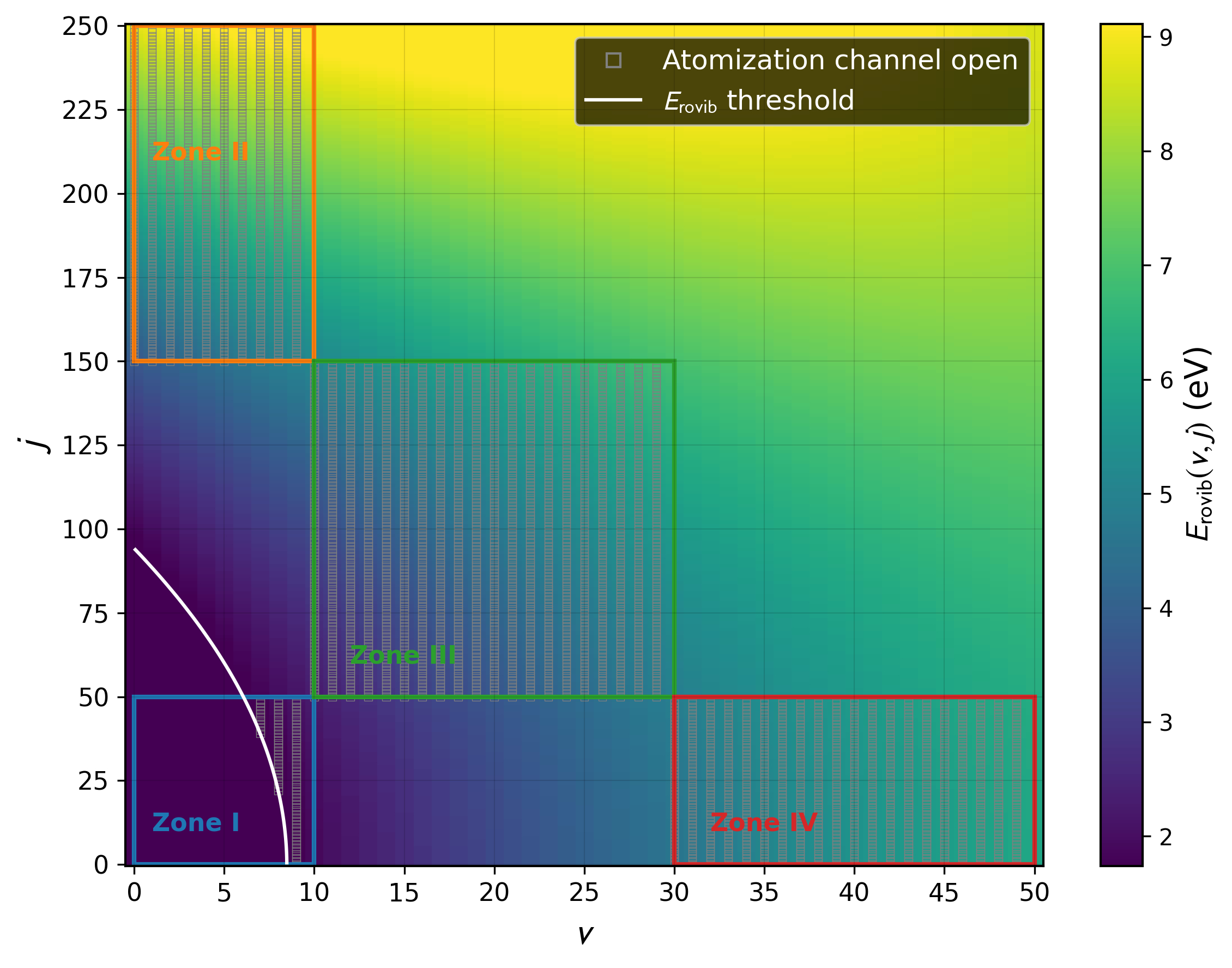}
	\caption{Rovibrational energy map
		$E_{\mathrm{rovib}}\mathrm{(}v,j\mathrm{)}$ for all initial
		states, with Zones I–IV indicated. Hollow squares mark $(v,j)$
		combinations where atomization is energetically accessible, within each chosen Zone. The
		white curve shows the atomization threshold.}
	\label{sifig:fig1}
\end{figure}

\begin{figure}[h!]
	\centering
	\includegraphics[width=1.0\linewidth]{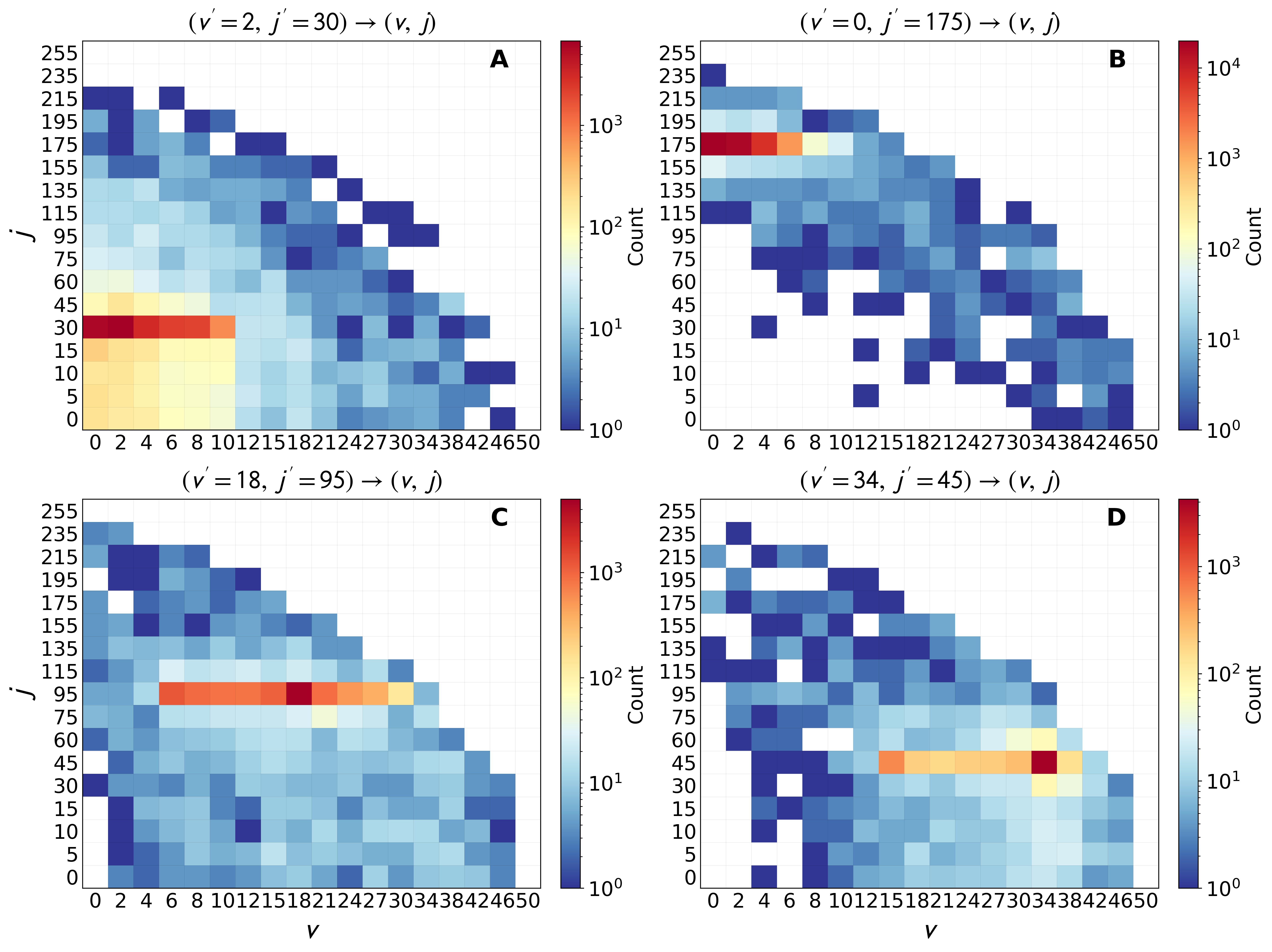}
	\caption{The N($^4$S)+O$_2$(X$^3 \Sigma_{\rm g}^+$) collision
		system. Origin maps of initial rovibrational states O$_2 (v,j)$
		that yield the indicated final states $(v',j')$ for the process
		N($^4$S) + O$_2$(X$^3\Sigma^-_{\rm g})$ $\leftrightarrow$
		NO$_2$.  Panels (top–left to bottom–right) correspond to
		$(v',j')=(2,30)$, $(0,175)$, $(18,95)$, and $(34,45)$.  Each
		panel shows a heat map of log-scaled counts of trajectories on
		the discrete grid of available initial states; white cells
		indicate no events. QCT simulations were carried out using the
		RKHS-PES.\cite{MM.no2:2020}}
	\label{sifig:fig2}
\end{figure}

\begin{figure}[h!]
	\centering
	\includegraphics[width=1.0\linewidth]{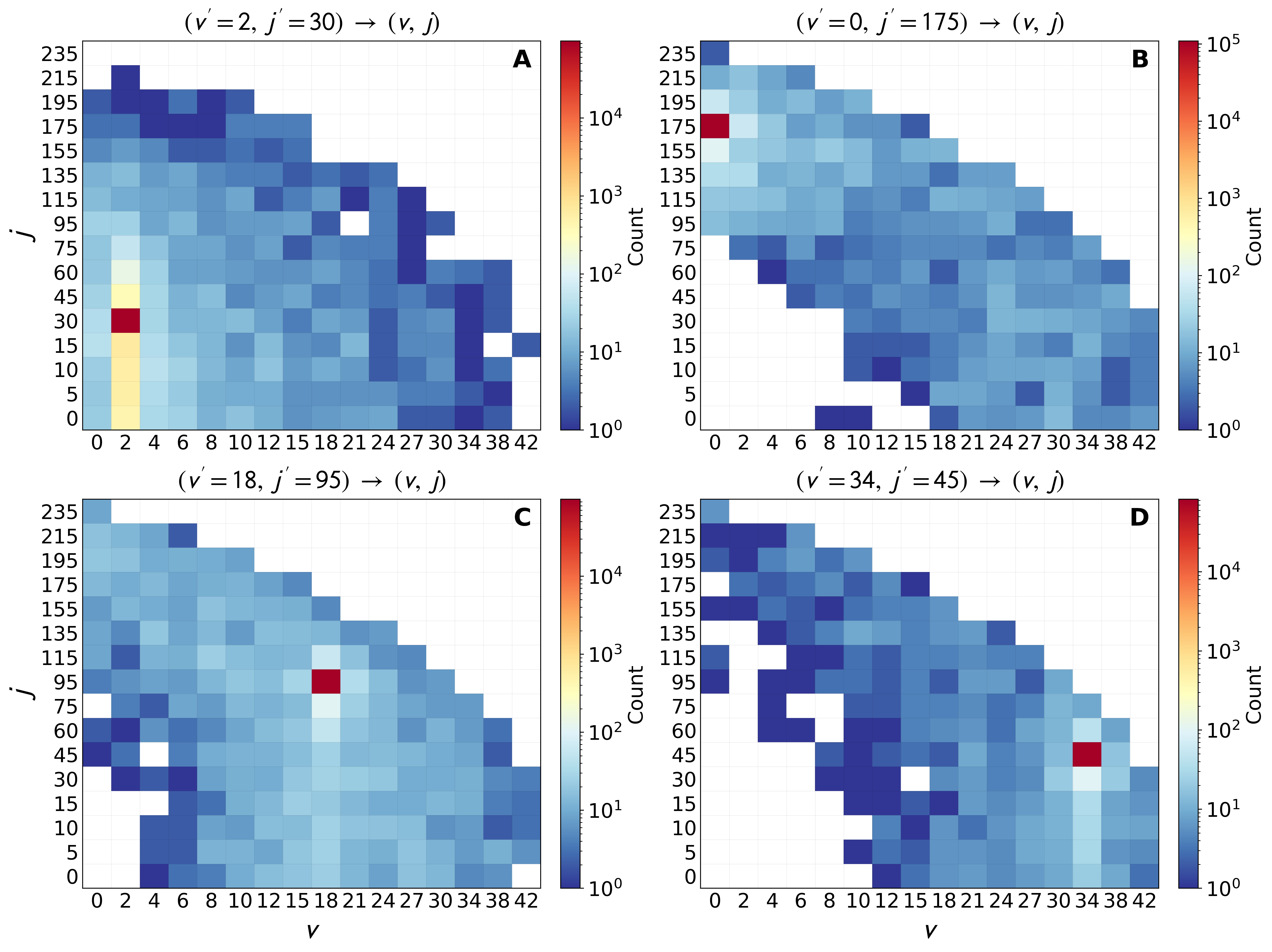}
	\caption{The N($^4$S) + NO(X$^2 \Pi$) collision system. Origin
		maps of initial rovibrational states NO(X$^2 \Pi) (v,j)$ that
		yield the indicated final states $(v',j')$ for the processes
		N($^4$S) + NO(X$^2 \Pi$) $\rightarrow$ O($^3$P) + N$_2$(X$^1
		\Sigma_{\rm g}^+$) (reaction) and N($^4$S) + NO(X$^2 \Pi$) $\rightarrow$
		N($^4$S) + NO(X$^2 \Pi$) (atom exchange). Panels (top–left to bottom–right) correspond to
		$(v',j')=(2,30)$, $(0,175)$, $(18,95)$, and $(34,45)$. Each
		panel shows a heat map of log-scaled counts of trajectories on
		the discrete grid of available initial states; white cells
		indicate no events. QCT simulations were carried out using the
		RKHS-PES.\cite{MM.n2o:2020}}
	\label{sifig:fig3}
\end{figure}

\end{document}